\documentclass[reprint,amsmath,amssymb, aps, prl]{revtex4-1}

\usepackage{graphicx} 
\usepackage{epsfig}
\usepackage{subfigure}
\usepackage{datetime}

\usepackage{bm}

\usepackage[usenames,dvipsnames]{color}
\usepackage[pagebackref=false, colorlinks=true]{hyperref}
\definecolor{redish}{rgb}{0.7,0.2,0.0}  
\definecolor{bluish}{rgb}{0.2,0.5,0.8}

\hypersetup{linkcolor=redish,     
                  citecolor=blue,        
                  filecolor=magenta,   
                  urlcolor=bluish}        

\begin{document}

\title{Mode stability of a near-extremal Kerr superspinar}

\author{Rittick Roy$^1$}
\email{rittickrr@gmail.com}
\author{Prashant Kocherlakota$^2$}
\email{k.prashant@tifr.res.in}
\author{Pankaj S. Joshi$^3$}
\email{psjprovost@charusat.ac.in}
\affiliation{$^1$Indian Institute of Technology, Mumbai 400076, India}
\affiliation{$^2$Tata Institute of Fundamental Research, Mumbai 400005, India}
\affiliation{$^3$International Center for Cosmology, Charusat University, Anand, GUJ 388421, India}

\date{\today}


\begin{abstract}
We study here the quasi-normal mode stability of a near-extremal Kerr superspinar, an exotic spinning compact object that exceeds the Kerr bound, under gravitational perturbations. Despite previous beliefs that these objects would be mode unstable, we show by analytically treating the Teukolsky equations that these objects are infact mode stable under almost all (barring a zero-measure set) boundary conditions.
\end{abstract}

\maketitle

\section{Introduction}
The spacetime geometry in the exterior of the asymptotic end state of a generic gravitational collapse for a rotating matter cloud is expected to be given by the Kerr solution of the Einstein equations \cite{Mars00}. The Kerr metric \cite{Kerr63} has conventionally been used to describe a rotating black hole or a naked singularity, depending on the angular momentum parameter $J$ and mass $M$ of the compact object.  When the Kerr bound $J \leq M^2$ is not violated, the final spacetime contains a Kerr black hole, that is, formation of an event horizon takes place, and in the case otherwise, a Kerr naked singularity occurs.

Properties of both classes of Kerr spacetimes, namely the black hole and naked singularity configurations, have been explored in great detail in various contexts such as, the structure of ergoregions \cite{Visser09, Chakraborty+17a} and shadows \cite{EHT19, Hioki_Maeda09}, efficiency of energy transfer from accretion disks \cite{Gimon_Horava09}, highly energetic particle collisions \cite{Banados+09}, gravitomagnetic spin-precession \cite{Chakraborty+17b}, to name a few. The overall picture that emerges is that if both Kerr black holes and Kerr naked singularities exist in nature, it might be possible to distinguish them from astrophysical observations. 

However, the question of whether these objects do, in fact, exist ubiquitously in nature falls within the ambit of a stability analysis of their corresponding exterior spacetimes. The quasi-normal mode frequencies (QNFs) of Kerr black holes were obtained after a vigorous study \cite{Teukolsky73, Teukolsky_Press74, Detweiler80, Leaver86}, and it was eventually established that Kerr black holes are indeed mode stable, in the seminal work of Whiting \cite{Whiting89}. On the other hand, the QNFs were also obtained recently for the Kerr naked singularity spacetimes \cite{Dotti+08}, and it is likely that the Kerr naked singularities may not be mode stable.

Now, given that the Kerr naked singularity spacetime is unstable against mode perturbations, does this mean that studies of its properties must be abandoned? Following \cite{Gimon_Horava09}, we adopt the following perspective. Singular metrics are solutions of the classical Einstein equations and the expectation is that a deeper theory of quantum gravity would smear out these singularities, irrespective of whether or not they are covered from asymptotic observers by event horizons. Therefore, it is possible that quantum gravity could introduce classes of legitimate compact objects such that their exterior geometries are described by metrics that were classically nakedly singular, but with their central singular regions excised and replaced by regions governed by Planckian physics. As argued by \cite{Gimon_Horava09}, string theory, a popular candidate for the quantum theory of gravity, has proven to be exceptionally good at resolving spacetime geometries with various timelike singularities, and such singularities in general relativity (GR) could in fact represent new classes of legitimate compact objects in the string-theoretic completion of GR (see for example the pair of papers, \cite{Breckenridge+97, Gimon_Horava04}.)

Therefore, following \cite{Gimon_Horava09}, we introduce the notion of a Kerr superspinar, a third exotic class of Kerr compact objects, whose exterior geometry is given by the overspinning Kerr metric ($J > M^2$). However, this hypothetical exotic object is assumed to have a finite size which, in the usual Boyer-Lindquist coordinates, is denoted by $r=r_0 > 0$. That is, the exterior spacetime of the Kerr superspinar is identical to that of the Kerr naked singularity, but its interior metric is treated as being unknown, and assumed to be provided by the full UV-complete theory of gravity.

An attempt to study the mode stability of these singularity-excised spacetimes was recently made \cite{Pani+10}. To better understand what this entails, it is useful to remember that the study of mode stability of any given spacetime requires us to solve the linearized Einstein equations with `appropriate boundary conditions.' Irrespective of the spacetime under consideration, one imposes the condition that there are no sources at asymptotic infinity, i.e. there is no incoming radiation at infinity. As for the boundary condition at the inner edge, in the case of black hole spacetimes, one naturally imposes perfectly absorbing boundary conditions at the horizon. However, for exotic objects like Kerr superspinars, there is no such `natural' boundary condition for incoming modes at its surface, and one must find quasinormal modes and their frequencies for each possible boundary condition. Pani \textit{et al} \cite{Pani+10} used a variety of boundary conditions that included all previous studies, and they found employing a numerical approach, that superspinars are typically likely to be unstable.

The equation governing the evolution of quasi-normal modes of the Kerr spacetime was discovered by Teukolsky \cite{Teukolsky73}. Recently, motivated by insight into the pole structure of the Teukolsky equation, we revisited the issue of mode stability of near-extremal ($J \gtrsim M^2$) superspinars in \cite{Nakao+18}. As a first step, we imposed boundary conditions identical to those that are imposed in the case of a black hole at its horizon, i.e. purely absorbing boundary conditions at the surface of the superspinar $r=r_0$. In this particular case, $r_0$ can be thought of as being the location of a `stringy horizon' of the Kerr superspinar. As an ansatz, we set the quasi-normal frequency spectrum of the near-extremal superpsinar to be identical to the near-extremal black hole and showed that for these boundary conditions, this spectrum is allowed. This ansatz was backed by the results of the numerical studies in \cite{Pani+10}. There it was discussed that near-extremal superspinars with stringy horizons are indeed mode stable. Therefore, the overspinning Kerr geometry could possibly admit legitimate interior solutions and must hence not be discarded without further exploration. 

In the current work, our purpose is to maximally extend our results to include arbitrary boundary conditions, imposed at the surface of the superspinar. We find that barring a zero-measure set of boundary conditions, corresponding to `almost perfectly reflecting boundary conditions', near-extremal superspinars are in fact generically mode stable.

\section{The Teukolsky Equation}
The Kerr metric in Boyer-Lindquist (BL) coordinates $(t, r, \theta, \phi)$ is given as,
\begin{align} \label{eq:KerrMetricBL} 
ds^2 =& -\left(1- \frac{2 M r}{\rho^2}\right)dt^2 -  \frac{4Mar\sin^2\theta}{\rho^2}dtd\phi \\
& \quad\quad + \frac{A\sin^2\theta}{\rho^2} d\phi^2  + \frac{\rho^2}{\Delta}dr^2 + \rho^2 d\theta^2 \nonumber, 
\end{align}
where we have employed geometrized units $G \! = \! c \! = \!1$. In the above, we have introduced the specific angular momentum $a = Ma_* = J/M$, with $\Delta=r^2-2Mr+a^2$, $\rho^2=r^2+a^2 \cos^2\theta$ and $A = (r^2 + a^2)^2 -a^2 \Delta\sin^2\theta$. 

The discovery of a single master equation that determined the evolution of separable perturbations of various types of fields (scalar, electromagnetic and gravitational), with a spin-weight parameter $s$ characterizing the type of the field was reported in a seminal paper by Teukolsky \cite{Teukolsky73}. There, it was argued that with the introduction of the Kinnersley complex null tetrad \cite{Kinnersley69}, the electromagnetic field can be characterized by its Newman-Penrose \cite{Newman_Penrose62} components. Further, it was identified that the $\phi_0$ and $\phi_2$ components correspond to the ingoing and outgoing radiative parts of the field. Similarly, gravitational radiation is described by perturbations in the Weyl tensor $C_{\alpha\beta\gamma\delta}$, the traceless component of the Riemann tensor, and has ingoing and outgoing radiative parts given by $\psi_0$ and $\psi_4$. These quantities are well-behaved invariants under gauge transformations and infinitesimal tetrad rotations. Teukolsky unified the perturbation treatment of all types of perturbative fields (scalar, electromagnetic, gravitational) by introducing a master variable,
\begin{equation}
\psi=e^{-i\omega t+im\varphi}R_{lm}(r)S_{lm}(\theta),
\end{equation}
where $\omega$ will acquire the interpretation of being the characteristic quasi-normal frequency (QNF), when the appropriate boundary conditions are imposed and is, in general, complex. As mentioned already, the appropriate boundary conditions for the above perturbations to be treated as quasi-normal modes are that there be no incoming waves at spatial infinity. Imposing this condition, we investigate here the QNFs of outgoing gravitational field perturbations, corresponding to $\psi_4$. In this case, we can write $\psi_4=(r-ia\cos\theta)^{-4}~\psi$, as can be seen from Table 1 of \cite{Teukolsky73}, for $s=-2$.

The governing linearized Einstein equations of motion for this class of perturbations $\psi$ are called the Teukolsky equations, and in terms of the radial $R_{lm}$ and angular $S_{lm}$ functions, are given as,
\begin{align}
&\Delta^{-s}\frac{d}{dr}\left(\Delta^{s+1}\frac{dR_{lm}}{dr}\right) \label{eq:R-eq} \\
& \quad\quad +\left(\frac{K^2-2is(r-M)K}{\Delta}+4is\omega r-\lambda\right)R_{lm}=0, \nonumber \\
&\frac{1}{\sin\theta}\frac{d}{d\theta}\left(\sin\theta\frac{dS_{lm}}{d\theta}\right) + \biggl[\left(a\omega\cos\theta+s\right)^2  \label{eq:S-eq}\\
&\quad\quad -\left(\frac{m+s\cos\theta}{\sin\theta}\right)^2-s(s-1)+F\biggr]S_{lm}=0, \nonumber
\end{align}
where $|s|$ denotes the spin of the perturbing field, i.e., $|s| = 0, 1, 2$ for scalar, electromagnetic and gravitational ($|s|=2$) perturbations. $F={}_sF^l_{m,\omega}$ with the integer $l$ larger than or equal to ${\rm max}(|m|,|s|)$ is the separation constant equivalent to the eigenvalue of (\ref{eq:S-eq}) with the boundary conditions of regularity at $\theta=0$ and $\pi$, $K:=(r^2+a^2)\omega-am$ and $\lambda:=F+a^2\omega^2-2am\omega$.

In the case of $a^2<M^2$, $r=r_\pm:=M\pm\sqrt{M^2-a^2}$ are real roots of $\Delta=0$; $r=r_+$ corresponds to the event horizon and $r=r_-$ is the location of the Cauchy horizon. In the extremal case, $a^2=M^2$, $r_+$ and $r_-$ agree with each other, and there is only one degenerate event horizon. In the case of $a^2>M^2$, i.e., a naked singularity or a superspinar, there is no real root of $\Delta=0$, and correspondingly no event horizon exists.

Now, following \cite{Cardoso04}, we introduce,
\begin{equation}
R_{lm} = \Delta^{-s}\tilde{R} \exp\left(-i\int\frac{K}{\Delta}dr\right),
\end{equation}
so that Eq.~(\ref{eq:R-eq}) becomes
\begin{align}
&\Delta\frac{d^2\tilde{R}}{dr^2}-\left[2i\omega (r^2+a^2)-2(\tilde{s}+1)(r-M)-2iam\right] \nonumber\\ 
&\times\frac{d\tilde{R}}{dr} -\left[2(2\tilde{s}+1)i\omega r+\tilde{\lambda}\right]\tilde{R}=0, \label{eq:tR-eq}
\end{align}
where, using $F=E-s(s+1)$, we have introduced 
$\tilde{s}=-s$ and $\tilde{\lambda}=\lambda+2s=E+a^2\omega^2-2am\omega-\tilde{s}(\tilde{s}+1)$.

\section{Quasi-Normal Stability of Near Extremal Superspinars}
We consider a near-extremal Kerr spacetime and hence we write the Kerr parameter in the form $$ a=M(1-\epsilon), $$ assuming $0<|\epsilon|\ll1$. The spacetime contains a superspinar in the case of $\epsilon<0$, 
whereas there is a black hole in the case of $\epsilon> 0$. 

In the case of black hole, it is known that the quasi-normal mode (QNM) frequency $\omega$ approaches $m/2M$ for  $m=l$ in the limit of $\epsilon\rightarrow0_+$ \cite{Detweiler80}. The numerical study in Ref.~\cite{Pani+10} has revealed that even in the superspinar case, the QNM frequency $\omega$ approaches $m/2M$ for $m=l$ modes in the limit $\epsilon\rightarrow0_-$. Hence, hereafter we focus on the modes of $m=l$ and assume 
\begin{equation}
M\omega-\frac{m}{2}={\cal O}\left(|\epsilon|^p\right), \label{assumption}
\end{equation}
where $p$ is a positive constant. 

We rewrite Eq.~(\ref{eq:tR-eq}) in terms of the the dimensionless variables $y:=(r-M)/M$ and $\tilde{\omega}:=M\omega$ as, 
\begin{align}
(y^2&-2\epsilon+\epsilon^2)\frac{d^2\tilde{R}_{lm}}{dy^2} -\left[2i\tilde{\omega} y^2+2(2i\tilde{\omega}-\tilde{s}-1)y \right. \nonumber\\
& \left. +2i(2\tilde{\omega}-m)(1-\epsilon) + 2i\tilde{\omega}\epsilon^2\right]\frac{d\tilde{R}_{lm}}{dy} \nonumber \\
& -\left[ 2(2\tilde{s}+1)i\tilde{\omega}(y+1)+\tilde{\lambda}\right]\tilde{R}_{lm}=0.
\label{Basic-eq}
\end{align}
We now divide this equation in two regions, the far zone defined as $y \gg max[\sqrt{|\epsilon|},|\epsilon|^p]$ and the near zone defined as $y \ll 1$. The solution to Eq.~(\ref{eq:tR-eq}) in the far zone  is written in terms of confluent hypergeometric functions $_1F_1(\alpha;\gamma;z)$; 
\begin{align}
\tilde{R}_{lm}^{~\text{far}}
&= A y^{-\tilde{s}-1/2+2i\tilde{\omega}+i\delta} \nonumber\\
&\times{}_1F_1\left(\frac{1}{2}+\tilde{s}+2i\tilde{\omega}+i\delta;1+2i\delta;2i\tilde{\omega} y\right)\nonumber \\
&+B y^{-\tilde{s}-1/2+2i\tilde{\omega}-i\delta}\nonumber\\
&\times{}_1F_1\left(\frac{1}{2}+\tilde{s}+2i\tilde{\omega}-i\delta;1-2i\delta;2i\tilde{\omega} y\right),
\label{solution-far}
\end{align}
where $A$ and $B$ are integration constants, and 
$$
\delta^2:=4\tilde{\omega}^2-\frac{1}{4}-\tilde{\lambda}-\tilde{s}(\tilde{s}+1)\simeq \frac{1}{4}(7m^2-1)-E.
$$ is a constant. For the near-zone analysis, we keep terms only of leading order in $\epsilon$ and introduce a new radial variable, $x:=y-\sqrt{2\epsilon}$. The solution of Eq.~(\ref{eq:tR-eq}) in the near-zone is expressed by using  
Gauss's hypergeometric function ${}_2F_1(\alpha,\beta;\gamma;z)$ in the form 
\begin{align}
\tilde{R}_{lm}^{~\text{near}} =& C~x^{-\tilde{s}+4i\tau/\sigma} {}_2F_1(1/2-2i\tilde{\omega}+i\delta+4i\tau/\sigma, \nonumber \\
& 1/2-2i\tilde{\omega}-i\delta+4i\tau/\sigma; 1-\tilde{s}+4i\tau/\sigma;-x/\sigma) \nonumber \\
& +D~ {}_2F_1(1/2+\tilde{s}-2i\tilde{\omega}+i\delta,1/2+\tilde{s}-2i\tilde{\omega}-i\delta;\nonumber \\
&1+\tilde{s}-4i\tau/\sigma;-x/\sigma), \label{solution-near}
\end{align}
where 
\begin{align}
\sigma :=& 2\sqrt{2\epsilon}~~~{\rm and}~~~
\tau :=& (1+\sqrt{2\epsilon})\tilde{\omega}-\frac{m}{2},
\end{align}
and $C$ and $D$ are integration constants. As shown in \cite{Cardoso04}, the boundary condition for purely ingoing waves at the black hole boundary is given by $R^{near}_{lm}=1$. If we compare our case with a black hole, then the term with $C$ blows up for points close to the horizon. Thus ($C=0$, $D=1$) is the boundary condition for purely ingoing waves and hence $C$ must be the reflection and $D$ the transmission coefficient.
Both solutions (\ref{solution-far}) and (\ref{solution-near}) are valid in the over-lapping region.  
In the limit $y\rightarrow0$, the solution (\ref{solution-far}) behaves as
$
\tilde{R}_{lm}\rightarrow Ay^{-\tilde{s}-1/2+2i\tilde{\omega}+i\delta} +By^{-\tilde{s}-1/2+2i\tilde{\omega}-i\delta}.
$
In the limit $y\rightarrow \infty$, the solution (\ref{solution-near}) behaves as
$
\tilde{R}_{lm}\rightarrow {\cal A}y^{-\tilde{s}-1/2+2i\tilde{\omega}+i\delta}+{\cal B} y^{-\tilde{s}-1/2+2i\tilde{\omega}-i\delta},
$
where $\cal A$ and $\cal B$ are given by
\begin{align}
&{\cal A}=\sigma^{1/2-2i\tilde{\omega}-i\delta}\Gamma(2i\delta) \nonumber \\
&\times\Biggl(
\frac{C\sigma^{4i\tau/\sigma}\Gamma(1-\tilde{s}+4i\tau/\sigma)}
{\Gamma(1/2-\tilde{s}+2i\tilde{\omega}+i\delta)\Gamma(1/2-2i\tilde{\omega}+i\delta+4i\tau/\sigma)} \nonumber \\
&+\frac{D\sigma^{\tilde{s}}\Gamma(1+\tilde{s}-4i\tau/\sigma)}
{\Gamma(1/2+\tilde{s}-2i\tilde{\omega}+i\delta)\Gamma(1/2+2i\tilde{\omega}+i\delta-4i\tau/\sigma)} 
\Biggr), \nonumber\\
 \label{calA}\\
&{\cal B}={\cal A}|_{\delta\rightarrow-\delta}. \label{calB}
\end{align}
Thus, equating the two solutions in the overlapping region we get
\begin{equation}
A={\cal A}~~~~{\rm and}~~~~B={\cal B}. \label{matching}
\end{equation}
From the far-zone solution (\ref{solution-far}), for $y\rightarrow\infty$, we have
\begin{equation}
\tilde{R}_{lm}^{\ \rm far}\simeq Z_{\rm out}~ y^{-(1-4i\tilde{\omega})}e^{2i\tilde{\omega} y} +Z_{\rm in}~ y^{-(2\tilde{s}+1)},\nonumber 
\end{equation}
where
\begin{align}
Z_{\rm in}&=A\frac{(-2i\tilde{\omega})^{-1/2-\tilde{s}-2i\tilde{\omega}-i\delta}\Gamma(1+2i\delta)}
{\Gamma(1/2-\tilde{s}-2i\tilde{\omega}+i\delta)} \cr
&+B\frac{(-2i\tilde{\omega})^{-1/2-\tilde{s}-2i\tilde{\omega}+i\delta}\Gamma(1-2i\delta)}{\Gamma(1/2-\tilde{s}-2i\tilde{\omega}-i\delta)}, \cr
Z_{\rm out}&=Z_{\rm in}|_{\tilde{s}\rightarrow-\tilde{s},\tilde{\omega}\rightarrow-\tilde{\omega}}. \nonumber
\end{align}
Thus, together with Eq.~(\ref{matching}), 
the no-incoming wave condition, $Z_{\rm in}=0$, leads to 
\begin{equation}
{\cal A}\frac{(-2i\tilde{\omega})^{-i\delta}\Gamma(1+2i\delta)}{\Gamma(1/2-\tilde{s}-2i\tilde{\omega}+i\delta)} 
+\left(\delta\rightarrow-\delta\right)=0
\label{QNM-condition}
\end{equation}
Eq. (\ref{QNM-condition}) along with Eq. (\ref{calA}) and (\ref{calB}) gives us the expression of the QNM frequencies in term of the reflection and transmission coefficient at the Superspinar boundary. Substituting Eq. (\ref{calA}) and Eq. (\ref{calB})  into Eq. (\ref{QNM-condition}) we get  
\begin{equation} \label{CD}
    D\sigma^{\tilde{s}}(f_2+\sigma^{2i\delta}f_4)=-C\sigma^{4i\tau/\sigma}(f_1+\sigma^{2i\delta}f_3)
\end{equation}
where $f_1, f_2, f_3, f_4$ are given by
\begin{align}{
f_1=&\Bigg(\frac{\Gamma(2i\delta)\Gamma(1+2i\delta)(-2i\tilde{\omega})^{-i\delta}}{\Gamma(1/2-\tilde{s}+2i\tilde{\omega}+i\delta)\Gamma(1/2-\tilde{s}-2i\tilde{\omega}+i\delta)}\Bigg)\nonumber\\
&\times\frac{\Gamma(1-\tilde{s}+4i\tau/\sigma)}{\Gamma(1/2-2i\tilde{\omega}+i\delta+4i\tau/\sigma)}\\
f_2=&\Bigg(\frac{\Gamma(2i\delta)\Gamma(1+2i\delta)(-2i\tilde{\omega})^{-i\delta}}{\Gamma(1/2+\tilde{s}-2i\tilde{\omega}+i\delta)\Gamma(1/2-\tilde{s}-2i\tilde{\omega}+i\delta)}\Bigg)\nonumber\\
& \times\frac{\Gamma(1+\tilde{s}-4i\tau/\sigma)}{\Gamma(1/2+2i\tilde{\omega}+i\delta-4i\tau/\sigma)}\\
f_3=&\Bigg(\frac{\Gamma(-2i\delta)\Gamma(1-2i\delta)(-2i\tilde{\omega})^{i\delta}}{\Gamma(1/2-\tilde{s}+2i\tilde{\omega}-i\delta)\Gamma(1/2-\tilde{s}-2i\tilde{\omega}-i\delta)}\Bigg)\nonumber\\
& \times\frac{\Gamma(1-\tilde{s}+4i\tau/\sigma)}{\Gamma(1/2-2i\tilde{\omega}-i\delta+4i\tau/\sigma)}\\
f_4=&\Bigg(\frac{\Gamma(-2i\delta)\Gamma(1-2i\delta)(-2i\tilde{\omega})^{i\delta}}{\Gamma(1/2+\tilde{s}-2i\tilde{\omega}-i\delta)\Gamma(1/2-\tilde{s}-2i\tilde{\omega}-i\delta)}\Bigg)\nonumber\\
& \times\frac{\Gamma(1+\tilde{s}-4i\tau/\sigma)}{\Gamma(1/2+2i\tilde{\omega}-i\delta-4i\tau/\sigma)}}\end{align}
With $C$ being the reflection coefficient and $D$ the transmission coefficient, we have $D^2=1-C^2$. In the regime $\sigma \to 0$, we can apply Stirling's formula to the above equations and simplify to get $f_1$, $f_2$, $f_3$ and $f_4$ respectively as 
\begin{eqnarray}
&\Bigg(&\frac{(4i\tau/\sigma)^{1/2-\tilde{s}+2i\tilde{\omega}-i\delta}\Gamma(2i\delta)\Gamma(1+2i\delta)(-2i\tilde{\omega})^{-i\delta}}{\Gamma(1/2-\tilde{s}+2i\tilde{\omega}+i\delta)\Gamma(1/2-\tilde{s}-2i\tilde{\omega}+i\delta)}\Bigg)\nonumber \\
&=&(4i\tau/\sigma)^{1/2-\tilde{s}+2i\tilde{\omega}-i\delta}g_1\\
&\Bigg(&\frac{(4i\tau/\sigma)^{1/2+\tilde{s}-2i\tilde{\omega}-i\delta}\Gamma(2i\delta)\Gamma(1+2i\delta)(-2i\tilde{\omega})^{-i\delta}}{\Gamma(1/2+\tilde{s}-2i\tilde{\omega}+i\delta)\Gamma(1/2-\tilde{s}-2i\tilde{\omega}+i\delta)}\Bigg)\nonumber \\
&=&(4i\tau/\sigma)^{1/2+\tilde{s}-2i\tilde{\omega}-i\delta}g_2
\end{eqnarray}
\begin{eqnarray}
&\Bigg(&\frac{(4i\tau/\sigma)^{1/2-\tilde{s}+2i\tilde{\omega}+i\delta}\Gamma(-2i\delta)\Gamma(1-2i\delta)(-2i\tilde{\omega})^{i\delta}}{\Gamma(1/2-\tilde{s}+2i\tilde{\omega}-i\delta)\Gamma(1/2-\tilde{s}-2i\tilde{\omega}-i\delta)}\Bigg)\nonumber \\
&=&(4i\tau/\sigma)^{1/2-\tilde{s}+2i\tilde{\omega}+i\delta}g_3\\
&\Bigg(&\frac{(4i\tau/\sigma)^{1/2+\tilde{s}-2i\tilde{\omega}+i\delta}\Gamma(-2i\delta)\Gamma(1-2i\delta)(-2i\tilde{\omega})^{i\delta}}{\Gamma(1/2+\tilde{s}-2i\tilde{\omega}-i\delta)\Gamma(1/2-\tilde{s}-2i\tilde{\omega}-i\delta)}\Bigg)\nonumber \\
&=&(4i\tau/\sigma)^{1/2+\tilde{s}-2i\tilde{\omega}+i\delta}g_4
\end{eqnarray}
Putting this back in Eq. (\ref{CD}),
\begin{equation}\label{25}
    \frac{C}{\sqrt{1-C^2}}=A\Bigg(\frac{g_2+(4i\tau)^{2i\delta}g_4}{g_1+(4i\tau)^{2i\delta}g_3}\Bigg)
\end{equation}
where we have defined
\[A=-\sigma^{\tilde{s}-4i\tau/\sigma} (-1)^{1/2+\tilde{s}-2i\tilde{\omega}-i\delta} (4i\tau/\sigma)^{2(\tilde{s}-2i\tilde{\omega})}\].
Note that the particular use of casting the equation in this form is to shift all the dependence of $\sigma$ to the term $A$. Writing $\tilde{\omega}=a+ib$ and simplifying $A$ we get
\begin{align}
A&=(4\tau/\sigma)^{\tilde{s}+2b+\sqrt{2}\sin{(1/2-p)}|\epsilon| ^{p-1/2}}\nonumber \\
&\times [(-1)^{3/2+3\tilde{s}/2-2ia+4b-i\delta}(4i\tau)^{\tilde{s}-4i\tau/\sigma}\nonumber \\
&\times (4i\tau/\sigma)^{\sqrt{2}i\cos{(1/2-p)}\pi|\epsilon| ^{1/2-p}-2ia}]
\end{align}
Keeping consistent with the argument presented in \cite{Nakao+18}, we would take $p<1/2$ from here onwards. It should be noted that in the above equation, the term inside $[...]$ is finite, but the term outside $(4\tau/\sigma)^{\tilde{s}+2b+\sqrt{2}\sin{(1/2-p)}|\epsilon| ^{1/2-p}}\rightarrow \infty$ since for $p<1/2$,  $(4\tau/\sigma)\rightarrow \infty$ and $|\epsilon| ^{p-1/2}\rightarrow \infty$ in the limit $\epsilon\rightarrow 0_\pm$ and $\sigma \rightarrow 0$. Simplifying Eq. (\ref{25}) by substituting $g_1, g_2, g_3, g_4$, we get
\begin{equation}\label{27}
    \Bigg(\frac{(-8\tilde{\omega} \tau)^{2i\delta}-q_2e^{i\chi_2}}{(-8\tilde{\omega} \tau)^{2i\delta}-q_1e^{i\chi_1}}\Bigg)=A'\frac{\sqrt{1-C^2}}{C}
\end{equation}
where we have defined
\begin{align}-q_1e^{i\chi_1}=&\frac{\Gamma(1/2+\tilde{s}-2i\tilde{\omega}-i\delta)\Gamma(1/2-\tilde{s}-2i\tilde{\omega}-i\delta)}{\Gamma(1/2+\tilde{s}-2i\tilde{\omega}+i\delta)\Gamma(1/2-\tilde{s}-2i\tilde{\omega}+i\delta)}\nonumber \\
&\times \frac{\Gamma(2i\delta)\Gamma(1+2i\delta)}{\Gamma(-2i\delta)\Gamma(1-2i\delta)}\end{align}
\begin{align}-q_2e^{i\chi_2}=&\frac{\Gamma(1/2-\tilde{s}+2i\tilde{\omega}-i\delta)\Gamma(1/2-\tilde{s}-2i\tilde{\omega}-i\delta)}{\Gamma(1/2-\tilde{s}+2i\tilde{\omega}+i\delta)\Gamma(1/2-\tilde{s}-2i\tilde{\omega}+i\delta)}\nonumber \\
&\times \frac{\Gamma(2i\delta)\Gamma(1+2i\delta)}{\Gamma(-2i\delta)\Gamma(1-2i\delta)}\end{align}
and
\[A'=A\Bigg(\frac{\Gamma(1/2-\tilde{s}+2i\tilde{\omega}-i\delta)\Gamma(1/2-\tilde{s}-2i\tilde{\omega}-i\delta)}{\Gamma(1/2+\tilde{s}-2i\tilde{\omega}-i\delta)\Gamma(1/2-\tilde{s}-2i\tilde{\omega}-i\delta)}\Bigg)\]
Taking dividendo of Eq. (\ref{27}) and rearranging we get
\begin{equation}
    (-8\tilde{\omega} \tau)^{2i\delta}-q_3q_1e^{i(\chi_1+\chi_3)}=0
\end{equation}
where 
\begin{equation}\label{31}
    q_3e^{i\chi_3}=1+\Bigg(\frac{C}{A'\sqrt{1-C^2}-C}\Bigg)\Bigg(1-\frac{q_2}{q_1}e^{i(\chi_2-\chi_1)}\Bigg)
\end{equation}
Following a similar analysis as \cite{Cardoso04}, we show that by taking $-8\omega \tau=\rho e^{i\zeta}$, the solution to the above equation is given by 
\begin{equation}
    \rho=e^{(\chi_1+\chi_3-2n\pi)/2\delta}
\end{equation}
\begin{equation}
    \zeta=-\frac{1}{2\delta}\log{(q_3q_1)}
\end{equation}
Since we are working in the limit $\omega \sim m\omega_+$, we can rewrite $-8\tilde\omega \tau$ as  $-8\tilde{\omega}\tau=-4m(\tilde{\omega}-m/2)$ , to obtain \cite{Sasaki_Nakamura90}
\[\tilde{\omega}=-\frac{\rho}{4m}\cos{\zeta}+m/2-i\frac{\rho}{4m}\sin{\zeta}\]
Therefore it follows that the stability  of the superspinar depends on the sign of $\sin{\zeta}$. It is evident from Eq. (\ref{31}) that as $A\rightarrow \infty$, $q_3\to 1$. It has already been shown in \cite{Nakao+18} that for $\zeta=-\frac{1}{2\delta}\log{(q_1)}$, $\sin{\zeta}>0$. If $q_3\to 1$, $\log{(q_3q_1)}\to$ $\log{(q_1)}$ and $\zeta\to-\frac{1}{2\delta}\log{(q_1)}$, and thus $\sin{\zeta}>0$, irrespective of what $C$ is. Thus, we conclude that the imaginary part of the QNM frequencies is always negative, except for the case $C=1$. This is because our analysis stems from the limiting behaviour of $\sigma$ in Eq. (\ref{31}) and in the case of $C=1$, $(1-C)$ goes to $0$ and hence the limiting behaviour of $\sigma$ does not affect the analysis. This case must be solved exactly and our analysis is not suited for  this case. It should also be noted here that for the case $C=0$, Eq. (\ref{31}) gives $q_3=1$ and hence stability is ensured as argued above, in agreement to the results of \cite{Nakao+18}. For the case $C=1$, \cite{Pani+10} has shown that the superspinar is unstable. Thus, in the entire parameter space of $C$, the extremal superspinar is stable for all boundary conditions except for the case $C=1$ and a zero-measure set very close to it.
\section{Conclusion}
In \cite{Nakao+18} the question raised was, are there boundary conditions for which the Superspinar is stable? In answer to this, it was suggested that we can take the frequency as the input parameter and set the imaginary part of it to be negative, and then find values of $C$ and $D$, the boundary parameters. Thus for a negative imaginary part of frequency there could exist an infinite set of $(C,D)$ for which the superspinar might be stable. In the present analysis, the new result that comes out is that, irrespective of what $C$ and $D$ are, a near extremal Kerr Superspinar would almost always be stable except for the $C=1$ case. The physical nature of the boundary corresponding to these boundary conditions is unclear at the moment and further study is required in this respect.

The present analysis is restricted to near extremal case for $l=m$ modes. But our result shows that for this particular case, the Kerr superspinar is almost always stable, in contrast to previous studies which concluded them to unstable. We do not exactly know yet what the boundary condition for the Superspinar at $r_0$ would be in physical reality. The interesting part of our result is that irrespective of what that boundary condition is, near extremal Kerr superspinars with $l=m$ modes would always be stable. With that being said, the statement in our earlier paper\cite{Nakao+18} is more strong since the argument could be extended to non-extremal case with $l \ne m$ modes. An interesting question then arises, namely, how strongly 
the properties of a classical nakedly singular metric, such as e.g. mode stability, shadows, images etc, would depend on the
existence of a central naked singularity itself, rather than there being a superspinar as the central exotic object. These issues will be discussed separately in a different paper.

\end{document}